\documentclass{WileyMSP-template}
\usepackage{chemformula} 
\usepackage[T1]{fontenc} 
\usepackage{comment}
\usepackage{upgreek}
\usepackage{textgreek}

\usepackage{lineno}

\begin{document}

\pagestyle{fancy}

\title{Bandgap Engineering On Demand in GaAsN Nanowires by Post-Growth Hydrogen Implantation}
\maketitle


\author{Nadine Denis}
\author{Akant Sharma}
\author{Elena Blundo}
\author{Francesca Santangeli}
\author{Paolo De Vincenzi}
\author{Riccardo Pallucchi}
\author{Mitsuki Yukimune}
\author{Alexander Vogel}
\author{Ilaria Zardo}
\author{Antonio Polimeni}
\author{Fumitaro Ishikawa}
\author{Marta De Luca*}

\begin{affiliations}
Nadine Denis, Alexander Vogel, Prof. Ilaria Zardo\\
Department of Physics, University of Basel, Basel, Switzerland\\

Dr. Akant Sharma, Dr. Elena Blundo, Francesca Santangeli, Paolo De Vincenzi, Riccardo Pallucchi, Prof. Antonio Polimeni, Prof. Marta De Luca\\
Dipartimento di Fisica, Sapienza Università di Roma, Rome, Italy\\
marta.deluca@uniroma1.it

Mitsuki Yukimune, Prof. Fumitaro Ishikawa\\
Graduate School of Science and Engineering, Ehime University, Ehime, Japan

Prof. Fumitaro Ishikawa\\
Research Center for Integrated Quantum Electronics, Hokkaido University, Sapporo, Japan

\end{affiliations}




\begin{abstract}
Bandgap engineering in semiconductors is required for the development of photonic and optoelectronic devices with optimized absorption and emission energies. This is usually achieved by changing the chemical or structural composition during growth or by dynamically applying strain. Here, the bandgap in GaAsN nanowires grown on Si is increased post-growth by up to 460 meV in a reversible, tunable, and non-destructive manner through H implantation. Such a bandgap tunability is unattained in epilayers and enabled by relaxed strain requirements in nanowire heterostructures, which enables N concentrations of up to 4.2\% in core-shell GaAs/GaAsN/GaAs nanowires resulting in a GaAsN bandgap as low as 0.97 eV. Using $\upmu$-photoluminescence measurements on individual nanowires, it is shown that the high bandgap energy of GaAs at 1.42 eV is restored by hydrogenation through formation of N-H complexes. By carefully optimizing the hydrogenation conditions, the photoluminescence efficiency increases by an order of magnitude. Moreover, by controlled thermal annealing, the large shift of the bandgap is not only made reversible, but also continuously tuned by breaking up N-H complexes in the hydrogenated GaAsN. Finally, local bandgap tuning by laser annealing is demonstrated, opening up new possibilities for developing novel, locally and energy-controlled quantum structures in GaAsN nanowires.

\end{abstract}

\section{Introduction}
III-V semiconductor nanowires (NWs) have a variety of applications for optoelectronic devices, including solar cells with enhanced light absorption \cite{Prete2020, Krogstrup2013}, photodetectors \cite{Li2023}, quantum \cite{Mäntynen2019, Versteegh2014} and classical light sources \cite{Yuan2021, Yi2022, Jansson2024}. The waveguiding properties of NWs allow to optimize the geometry to maximize absorption at desired wavelengths, or to achieve highly efficient photon emission and extraction properties \cite{Prete2020, Dalacu2019}. Ga(In)AsN NWs are particularly interesting for photonic applications as they have been shown to reach telecom wavelengths \cite{Nakama2023}, while being monolithically integrated on large-scale industry standard Si substrates \cite{Minehisa2024}.

The field of quantum structures in NWs encompasses a number of designs for efficient, high quality non-classical light emitters embedded in bottom-up grown NWs for application in quantum photonic technologies \cite{Dalacu2019, Mäntynen2019}. By advanced bandgap engineering, several quantum dots can potentially be grown in the same NW waveguide and phenomena such as multiplexing \cite{Laferriere2020} or superradiance \cite{Grim2019} can be investigated. Or, by utilizing the core-shell-shell structure of NWs, it is possible to study phenomena such as excitonic Ahranov-Bohm oscillations in NW quantum rings \cite{Corfdir2019, Royo2017}. An important advantage of the NW system is the possibility to grow them site-controlled or to pick them up and place them in a photonic circuit or cavity for further applications, a process that can be automated for large-scale applications \cite{Jevtics2020}.

However, the integration of quantum structures into NWs requires precise spatial control over the bandgap energy at the nanoscale. Tremendous progress has been made to develop strategies during the growth process, which typically involve changing the material composition, crystal phase or introducing strain to the material with a lattice mismatched outer shell \cite{Sun2024,Balaghi2019} or oxide layer \cite{Bavinck2012}. Alternatively, reversible, dynamic approaches have been explored to tune the bandgap after growth \cite{Greil2016, Signorello2014, Fiset2018}, but they require complex device designs.

In this work, we demonstrate that by implanting H post-growth into the GaAsN shell by exposing the NW to low-energy ionized H gas, we can continuously tune the bandgap by an unprecedented extent of about 0.5 eV. This is indeed attained thanks to the distinctive geometry of the NWs. By locally controlling the H implantation, this opens up a new way to create (multiple) size- and site-controlled quantum dots and quantum rings within a NW.

Indeed in GaAsN, the N atoms have energy states that are resonant with the states in the conduction band of GaAs, creating a perturbation potential that leads to the splitting of the previously degenerate conduction band into an upper and lower band as described by the band anti-crossing (BAC) model \cite{Shan1999}, which is discussed in more detail in SI 1. This downward bending of the lower conduction band leads to a redshift of the GaAsN bandgap with respect to GaAs and can be reversed by implanting H into the GaAsN lattice, where H binds to the N atoms and passivates their electronic potential, shifting the bandgap energy back to the value of GaAs \cite{Felici2006, Filippone2020}. Site-controlled H implantation in GaAsN quantum wells and epilayers was successfully used to develop quantum dots on demand \cite{Felici2020, Biccari2018, Birindelli2014}. However, for a long time, it was not possible to grow this promising material in the form of NWs, due to major challenges such as the low solubility of N in GaAs \cite{Stringfellow2021}.

Core-shell-shell GaAs/GaAsN/GaAs NWs with concentrations of N atoms up to 4.2\% exhibiting good optical emission up to room temperature (RT) were recently obtained \cite{Yukimune2018,Yukimune2019,Yukimune2020}. The incorporation of such high N concentrations while maintaining good optical material properties is possible in NWs owing to elastic strain deformation due to the large surface-to-volume ratio in NWs and to a distributed strain accommodation among the different material layers in the core-shell-shell NW heterostructure \cite{Glas2015, McIntyre2020, Balaghi2019}. Thus, it is now possible to fully exploit the advantages of NWs for the related optoelectronic and photonic applications, such as growth on Si and position control in combination with the hydrogenation process.

Here, we explore and optimize the hydrogenation of the GaAsN shell for tuning its bandgap on demand. Using \textmu-PL measurements at RT, we show a complete recovery of the GaAs bandgap after hydrogenation, which is accompanied by a considerable signal increase of one order of magnitude for optimized hydrogenation conditions. $\upmu$-PL scans of single GaAs/GaAsN NWs show a consistent hydrogenation along the entire NW, irrespective of the crystal phase and the presence of local defects. Through annealing of the hydrogenated NWs, the bandgap energy is controllably shifted over the range between the emission energy of GaAsN and GaAs, offering the possibility to engineer the absorption/emission energy of the nanostructure to the desired value. With local laser annealing, we further show how this technique can be used to tune the bandgap at a desired point of the NW. 
Hydrogenation of GaAs/GaAsN NWs can be directly utilized for matching energies in the design of NW photonic devices, not only for single photon based quantum devices, but also energy-matched detectors, spin filters, NW lasers or NW solar cells reaching the telecom O-band and potentially the telecom C-band.

\section{Results and discussion}
\subsection{Bandgap engineering of the GaAsN NW}
\begin{figure}
 \centering
 \includegraphics[width=\linewidth]{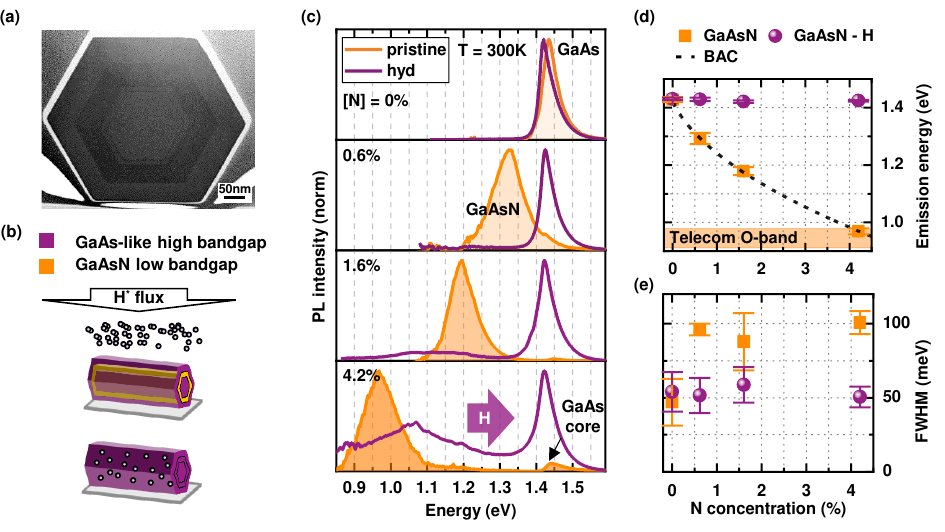}
 \caption{Bandgap engineering of the GaAsN NW shell. (a) Bright-field STEM image of the core-multishell GaAs/GaAsN/GaAs NW cross-section with 1.6\% N, a darker contrast is observed in the GaAsN shell compared to the GaAs core and outershell. (b) illustrates how the energy of the GaAsN bandgap (orange) shifts to the bandgap energy of GaAs (purple) upon H implantation of a single core/shell/shell GaAs/GaAsN/GaAs NW lying on a substrate. (c) shows RT \textmu-PL spectra measured in a point on single, transferred NWs before (shaded orange) and after (purple solid line) hydrogenation. The N content in the GaAsN shell of these NWs is 0, 0.6\%, 1.6\% and 4.2\% (spectra from top to bottom). After hydrogenation, and the passivation of N-atoms by H, the bandgap of all samples is upshifted to the value of GaAs, by 140 meV, 250 meV and 460 meV respectively. The H-dose here is $d_H = 1.2-1.4 H_0$, where $H_0$ corresponds to $10^{19}$ H-ions/cm$^2$. Normalization factors discussed in Figure \ref{fig02} and reported in the SI 2. Notice that the small post-hydrogenation redshift in the 0\% sample is not statistically relevant, as can be seen in the SI 2 and 4 and in panel (d) here. (d) and (e) show the energy at the maximum of the emission band and FWHM of the GaAsN shell before (orange squares) and after (purple dots) H implantation as a function of different N concentrations. The values are the average from a collection of several measurements on different NWs shown in SI 2. The error bars are given by the standard deviation. The dotted line in (d) shows the trend of the bandgap energy as a function of the N concentration according to the BAC and the area shaded orange shows the energy range of the telecommunications O-band. }
 \label{fig01}
\end{figure}

The NWs investigated in this work have a 170 nm thick GaAs core, a 40-50 nm thick GaAsN shell with N concentrations of 0.6\%, 1.6\% and 4.2\% and a 40-50 nm thick GaAs outer shell. GaAs NWs grown under the same conditions with 0\% N are used as reference. A bright-field scanning transmission electron microscope (STEM) image of the NW cross-section with 1.6\% N is shown in \textbf{Figure \ref{fig01}}(a). The NW growth is described in the methods and a detailed structural analysis can be found in \cite{Yukimune2019,Yukimune2020}.

First, we investigate the shift of the electronic bandgap of the GaAsN NW shell to lower energy with increasing N concentration and then the reversal of this shift after H implantation performed on single horizontal NWs, as illustrated in the schematic of Figure \ref{fig01}(b). For the experiment, the PL emission at RT is measured on several individual NWs in the same point before and after hydrogenation. The PL emission peak at RT is considered here as the bandgap, since impurity states are ionized.
Representative $\upmu$-PL spectra at RT before (shaded) and after (solid line) hydrogenation are shown in Figure \ref{fig01} (c) for increasing N concentration from top to bottom. In the pristine NWs, the bandgap emission of the GaAsN shell shifts to lower energy with increasing N concentration. After hydrogenation the PL spectra reveal a complete recovery of the higher GaAs-like bandgap energy.  In the samples with 1.6\% N and 4.2\% N, a low intensity emission band is visible at energies between 0.95-1.3 eV, it originates from H-activated radiative transitions between a H-donor level and the Ga-vacancy center (V$_{Ga}$-H) \cite{Capizzi1992a, Capizzi1992b, Chang1993,Bonapasta1992} as discussed in Figure \ref{fig02}. A quantitative analysis of the emission energy of the bandgap and the full width at half maximum (FWHM) upon hydrogenation is shown in Figure \ref{fig01} (d) and (e), respectively. These values are averages from a collection of measurements done on dozens of NWs shown in SI 2. In the pristine NWs, the bandgap energy in the GaAsN shell decreases from 1.43 eV for the NW with 0\% N to 1.29, 1.18 and 0.97 eV for NWs with 0.6, 1.6 and 4.2\% N respectively. The dashed line in Figure \ref{fig01}(d) marks the reduction of the bandgap according to BAC model that was used to determine the N-concentration of the NW samples (see SI 1). However, the true N concentration in the NW shell is expected to differ slightly, as the smaller lattice constant of the GaAsN material compared to pure GaAs causes some tensile strain in the GaAsN shell. The orange shadowed band in Figure \ref{fig01}(d) indicates the energy between 0.91 - 0.98 eV within the telecom O-band. After hydrogenation, the bandgap emission of all samples blueshifts to the same energy of the pure GaAs reference NWs. The microscopic origin of the recovery of the high GaAs bandgap lies in the passivation of the electronic potential that the N-atoms have on the GaAsN lattice through the formation of N-H complexes with at least two H atoms per N atom \cite{Filippone2020, Wen2012}.

The relatively broad and symmetric lineshape of the PL emission in untreated NWs is caused by N concentration fluctuations typical for GaAsN, due to a low solubility of N in GaAs \cite{Stringfellow2021, Buyanova2003}. These local concentration fluctuations disappear as soon as N is passivated by H, resulting in a linewidth narrowing by a factor 2 after hydrogenation (Figure \ref{fig01} (e)) and a restoring of a typical asymmetric PL-lineshape following the Boltzmann distribution at RT in all spectra (Figure \ref{fig01} (c)).

The emission of the GaAs core indicated in panel (c) has a very low intensity compared to the emission of the GaAsN shell, that is due to an efficient carrier transfer from the high bandgap GaAs material to the low bandgap GaAsN material, which is typical for core-shell NWs \cite{Chen2016, DeLuca2013}. Before hydrogenation, the emission energy of the GaAs core shows a blueshift (below 30 meV) for increasing N concentration in the GaAsN shell. This blueshift is related to the smaller lattice constant of GaAsN compared to GaAs, which leads to a compressive strain in the GaAs core causing a bandgap increase. Such a strain distribution through elastic deformation of both, the core and the shell, is typical for highly lattice mismatched core-shell NWs \cite{Balaghi2019, Glas2015, Gronqvist2009}. Interestingly, after hydrogenation the GaAsN emission of all samples shifts to the same GaAs-like energy due to the reversal of strain in the GaAsN shell through the formation of N-H complexes \cite{Geddo2011}.

The following sections systematically explore the precise control and understanding of phenomena accompanying the remarkable bandgap tuning through hydrogenation.
  
\subsection{Optimization of hydrogenation conditions}
H diffusion during hydrogenation, which is promoted by heating the sample, is strongly influenced by the trapping of H atoms in N-H complexes, altering typical diffusion models. Local electronic potentials, caused by N atoms or lattice/interface defects, act as H-traps, with the probability of trapping depending on the potential strength and the sample temperature. Sharp bandgap profiles  between passivated high bandgap GaAsN-H and the low bandgap GaAsN material are critical for precise quantum structure engineering. This is achieved when the H atoms are trapped by the first unpassivated N atom at relatively low hydrogenation temperatures below $T=250^\circ$C in epilayers \cite{Trotta2012, Trotta2011}, at higher hydrogenation temperatures the stronger H-diffusion leads to smoother transitions.

\begin{figure}
 \centering
 \includegraphics[height=6.5cm]{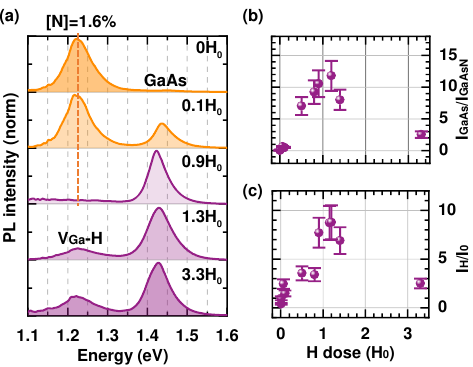}
 \caption{Effect of an increasing H-dose on the RT PL emission of the GaAs/GaAsN/GaAs core-shell NW with 1.6\% N. (a) shows \textmu-PL spectra of NWs hydrogenated with the indicated H-dose, where H$_0$ corresponds to $10^{19}$ H-ions/cm$^2$. The interplay between the emission intensity at low and high bandgap energy is measured as a ratio of the integrated PL intensity over the respective band for increasing H-doses and shown in (b). (c) shows the PL intensity increase at RT as a function of H-dose calculated as the ratio of the integrated emission intensity over all bands of the hydrogenated and the untreated NW. The data points in panels (b) and (c) are averages of at least 3 different points from different NWs and the error bars the uncertainties on the measured PL intensity.}
 \label{fig02}
\end{figure}

To systematically determine an optimal hydrogenation dose, NWs were measured in the same point before and after hydrogenation over a wide range of H doses. \textbf{Figure \ref{fig02}} (a) shows RT $\upmu$-PL spectra at the same point on a NW with 1.6\% N after successive hydrogenation. At low H dose, the intensity of the GaAsN band emitting at low energy (indicated by a dashed orange line) decreases and the emission intensity of the high-energy GaAs-like band at 1.43 eV increases as more and more N centers are passivated. Until, at an H dose of 0.9 H$_0$, only the high-energy GaAs-like bandgap emission remains. This direct transition from low to high energy bandgap emission confirms the sharp boundaries between passivated and an unpassivated GaAsN material. Instead, if the passivated N atoms were randomly distributed across the NW shell, we would observe an effective N concentration leading to a gradual and not a sudden blueshift of the GaAsN bandgap with increasing H dose.
At a high H dose, where all N atoms are passivated, a broad V$_{Ga}$-H emission band at energies between 1.1-1.3 eV is observed (as visible in the two bottommost spectra), which is attributed to a V$_{Ga}$-H complex. After supersaturation, the interstitial H that is introduced in the crystal activates a transition with Ga-vacancy centers (V$_{Ga}$-H) \cite{Capizzi1992a, Capizzi1992b, Chang1993}, whose broad character is explained by an overlap of more than one ground state of the V$_{Ga}$-H transition \cite{Bonapasta1992}. For the samples with 0 and 0.6\% N, we observe analogous hydrogenation dynamics (discussed in SI 3.1). 

The ratio between the maximum intensity of the GaAs-like and GaAsN-like emission for increasing H-dose is shown in Figure \ref{fig02}(b). We observe an increase with H-dose of the GaAs-like emission and then a decrease due to the appearance of the V$_{Ga}$-H transition at lower energy. The optimal H-dose is identified between 0.9 - 1.2 H$_0$ with H$_0 = 10^{19}$ H-ions/cm$^2$. Furthermore, we study the overall RT PL emission gain for an increasing H dose, which is calculated as the ratio between the total integrated PL intensity from the same point before and after hydrogenation. The average signal gain from several points is shown as a function of the H-dose in Figure \ref{fig02}(c), outliers with a difference of more than 2 standard deviations from average were discarded. The error bars in (b) and (c) reflect the experimental setup efficiency that can vary within 5\% depending on daily alignments and precise positioning on the center of the NWs. Significant differences in the intensity gain from point to point are observed related to a different crystal quality and structure in different points of the NW lattice. Individual points gain up to a factor 50 in intensity for samples with 1.6\% N. For an optimal H-dose, the mean intensity gain for NWs with 1.6\% N is a factor 8.5, for NWs with 0\% N it is 13. However, the NWs with low 0.6\% N show no change in PL emission intensity through hydrogenation, as shown in SI 3.1. The increase in luminescence efficiency can be accounted for by the passivation of the electrical activity and related optically detrimental effect of point defects, such as dangling bonds \cite{Dautremont1986, Pearton1987}. These defects are typically present at semiconductor interfaces, especially for lattice mismatched materials as it is the case for the GaAs/GaAsN system \cite{Buyanova2003}.
Finally, the strong attenuation of the total RT PL intensity, occurring in all the samples when being highly supersaturated with H, has also been related to the appearance of the V$_{Ga}$-H band \cite{Capizzi1992a, Chang1993}.

While good hydrogenation conditions were determined for GaAsN thin films, hydrogenation has never been attempted in any NW system and surface or diffusion dynamics might be different. The impact of the H beam energy and hydrogenation temperatures is analyzed in detail in SI 3.2 and 3.3 respectively. A systematic study for varying beam energy at a constant H-dose and temperature shows a more complete N-passivation when using an ion beam energy of at least 100 eV. However, in the NWs hydrogenated at beam energy higher than 120 eV, we observe a small increase of the FWHM by approximately 25\% coinciding with a 35\% decrease in optical emission gain. We conclude, that even though the employed ion beam energy is low compared to typical H-plasma treatments and avoids related damage effects, it is still beneficial to use the lowest energy possible that is large enough to overcome surface barriers and native oxide shells. In the case of the core-shell NW system studied here, we find an optimal ion beam energy of 100 eV.

\subsection{Full hydrogenation of the polytypic GaAsN NW}
\begin{figure*}
 \centering
 \includegraphics[width=0.95\linewidth]{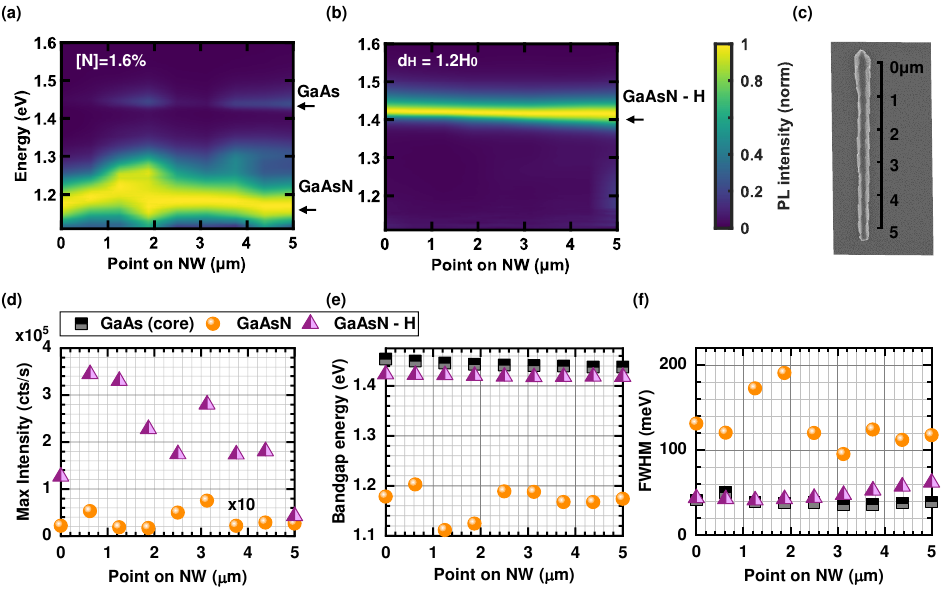}
 \caption{Bandgap tuning along a NW with 1.6\% N to a uniform GaAs-like energy in all regions of the NW. (a) and (b) show a map of RT \textmu-PL spectra as a function of the position along the axis of the same NW before and after hydrogenation. The emission energy of GaAs and GaAsN is marked by arrows. After hydrogenation only emission at GaAs-like energy is visible. All \textmu-PL spectra are individually normalized to 1 to highlight variations in the bandgap energy. The normalization factors are shown in (d) as the maximum PL intensity at the respective position measured in counts per second. An SEM image of the NW is shown in (c). Quantitative information regarding the maximum emission intensity, the local bandgap energy and FWHM along the NW axis is summarized in (d)-(f). For the pristine NWs the GaAs (black squares) and GaAsN-shell (orange circles) emission are analyzed separately. Data points along the hydrogenated NW are marked with purple triangles.}
 \label{fig03}
\end{figure*}
While numerous studies have demonstrated the shift of the GaAsN bandgap in the ZB crystal structure in epilayers, we reveal here that the same remarkable N-passivation through hydrogenation extends to WZ GaAsN. Previous TEM measurements of the samples investigated here have revealed phase changes showing about 80\% ZB structure and 20\% WZ structure \cite{Yukimune2019, Yukimune2020}. With RT $\upmu$-PL scan along the axis of single NWs, we show that a full shift of the GaAsN bandgap is achieved and H diffusion succeeds through a polytypic NW. The high carrier diffusion at RT ensures a probing of the entire NW cross-section, which proves the full hydrogenation of the NW.

\textbf{Figure \ref{fig03}}(a) shows a  RT $\upmu$-PL scan along the axis of a single core-shell-shell NW with 1.6\% N. The emission energy and peak width of the GaAsN shell varies depending on the position and the GaAs core emission is faintly visible at higher energy (see side arrows). However, despite these variations, we create a homogeneous sample after hydrogenation, as shown in the RT $\upmu$-PL scan in Figure \ref{fig03}(b). The SEM image of the measured NW in (c) shows some variations in diameter along the NW, which may impact the PL emission intensity along the NW. Each spectrum in (a) and (b) is normalized individually. The normalization factors are shown in Figure \ref{fig03}(d) as the maximum RT PL intensity in each point along the NW, with a point dependent intensity increase of approximately a factor 30 after hydrogenation.

The energy at the maximum of the RT PL emission peak and the corresponding FWHM of the GaAs core and GaAsN shell and the hydrogenated GaAs-like emission are shown as a function of position along the NW in (e) and (f). Changes in the RT PL emission peak in terms of energy and linewidth along the NW before hydrogenation can be due to different N incorporation due to the local crystal structure \cite{Li2016} and due to a modified bandgap bowing that depends on the distance between the energy level of the N atoms to the GaAs conduction band, which is different for WZ and ZB GaAs. After hydrogenation, the emission becomes uniform in terms of energy and linewidth along the entire NW (similar trends occur in the $\upmu$-PL scans along NWs with 0.6\% N as shown in the SI 4, where also a scan of the GaAs reference sample is reported). In general, the FWHM of GaAsN shell is smaller after hydrogenation, but slightly higher than the FWHM of the pristine GaAs core emission as already observed in Figure \ref{fig01}(e). The blueshift of the GaAs core due to compressive strain in the untreated NW is well visible in Figure \ref{fig03}(e).

We conclude, that even for a same N incorporation along a single NW, it is reasonable to expect a slightly different bandgap emission depending on the crystal structure. However, despite these variations in morphology and crystal phase, the sample is homogeneous in terms of bandgap energy and FWHM after hydrogenation, proving that a complete N passivation of the GaAsN shell is achieved.

\subsection{Tunability of the bandgap and reversal of H implantation by thermal annealing}
\begin{figure}
 \centering
 \includegraphics[]{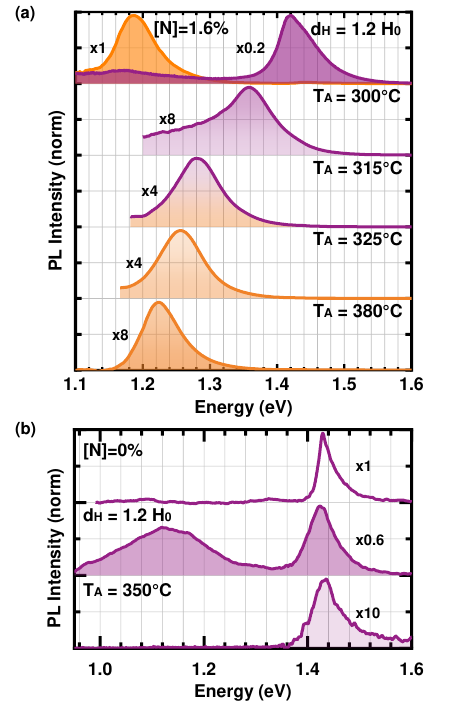}
 \caption{Tunability of the bandgap and reversal of H implantation by thermal annealing. (a) $\upmu$-PL spectra showing the bandgap emission in the same point on a hydrogenated NW after annealing at different temperatures as indicated. The N concentration is 1.6\%. The emission of the NW before (orange) and after (purple) hydrogenation are shown in the top row and the gradual recovery of the low energy bandgap of GaAsN after annealing at increasing temperatures in the spectra below. (b) Effects of implanted H on the pure GaAs reference samples. The \textmu-PL spectrum of a point on a pristine GaAs NW is shown in the top row, after hydrogenation in the second row and after mild annealing and the removing the of inserted H states in the bottom row. The spectra are normalized to the the maximum PL intensity of the pristine sample; normalization factors are given for all spectra and they reflect well those in Figure \ref{fig02} (c) and in the SI for both samples. At high annealing temperatures, a general decrease of signal is observed showing due to crystal quality deterioration.}
 \label{fig04}
\end{figure} 

In this section it is shown that thermal annealing of hydrogenated GaAsN NWs enables precise control over the GaAsN bandgap energy, tuning the emission from the hydrogenated GaAs-like energy at 1.43 eV to the low bandgap energy of GaAsN. \textbf{Figure \ref{fig04}}(a) shows $\upmu$-PL spectra at RT, measured at the same NW location before and after hydrogenation (top row) and following annealing at various temperatures (lower rows). Annealing at temperatures higher than the activation energy for dissolving certain N-H complexes reduces the effective concentration of unpassivated N, resulting in a gradual bandgap shift to higher energies, as evident in the spectra. Besides the continuous bandgap tuning itself, which is particularly relevant for energy matched applications, we stress that this result is pivotal, because it proves that the sudden blueshift of the GaAsN bandgap observed throughout this article after hydrogenation is not the result of the simultaneous quenching of GaAsN emission and enhancement of GaAs-core emission. Instead, it is due to the passivation of N atoms through the formation of N-H complexes, which is why a lower effective concentration of unpassivated N in the annealed NWs results in a bandgap intermediate between GaAs and the original GaAsN.

Figure \ref{fig04} (b) shows the RT $\upmu$-PL spectra of a spot on a GaAs NW measured in the untreated, highly hydrogenated and annealed state. This study on the N-free GaAs NWs is important because it allows to understand which signatures of H in the GaAs/GaAsN/GaAs NW lattice are unrelated to the passivation of N defects, but related to the presence of H atoms in the crystal lattice. Indeed, after thermal annealing and removal of the H donor state that is necessary for the radiative V$_{Ga}$-H transition, the emission band at energies between 1.0 and 1.3 eV disappears, as visible in the lowest spectrum in Figure \ref{fig04}(b). Furthermore, we observe that the redshift of the GaAs emission band and the increase in the FWHM upon hydrogenation are almost completely reversible by thermal annealing. Therefore, we argue that their occurrence is linked to the presence of H in the lattice of GaAs or GaAsN/GaAs NWs and not to other defects that could arise from ion bombardment.
\begin{figure}[h!]
 \centering
 \includegraphics[width=0.95\linewidth]{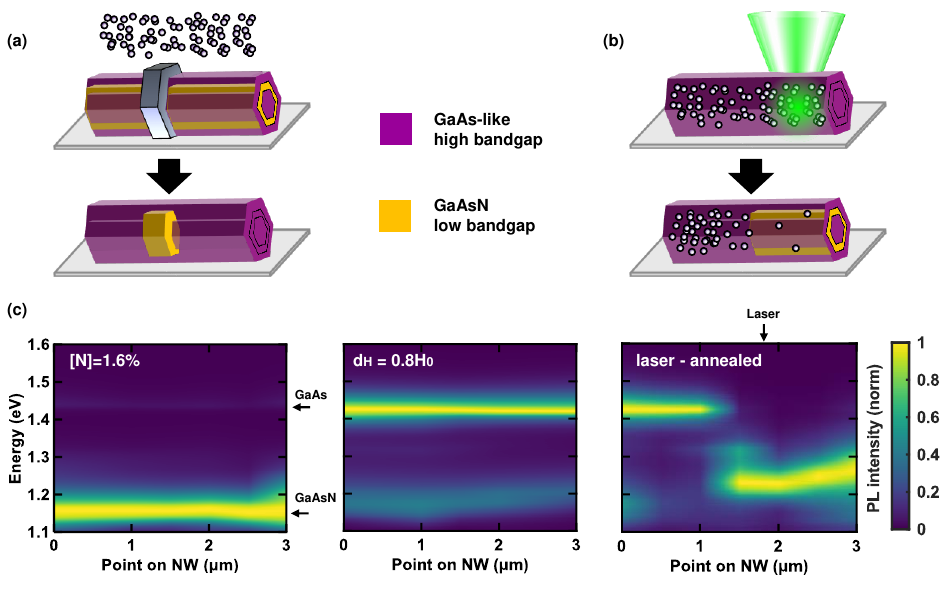}
 \caption{Two approaches for local bandgap engineering through hydrogenation. (a) illustrates the concept for achieving local bandgap tuning by a spatially selective hydrogenation approach, where part of the NW is masked by a H-opaque mask, below which GaAsN stays unpassivated. In the unmasked parts of the NW, the bandgap is shifted to the high GaAs-like energy (purple), confining the carriers in the low bandgap of the unpassivated GaAsN (orange). (b) shows a reverse approach, consisting in bandgap tuning by local laser annealing a hydrogenated high bandgap NW, locally removing the H below the laser spot, leading to low bandgap unpassivated GaAsN painted orange in the part illuminated by the laser beam. (c) shows a proof of concept for the second approach by laser annealing in the PL emission of a NW measured before hydrogenation, after hydrogenation and after laser annealing at coordinates of 2 $\upmu$m along the NW axis is shown from left to right. Owing to laser annealing, the bandgap is indeed shifted to low energy on one side of the NW, while the high energy bandgap of the hydrogenated GaAsN is conserved on the other side of the NW.}
 \label{fig05}
\end{figure}

\subsection{Local bandgap engineering post-growth}
Local bandgap engineering by post-growth hydrogenation can be used to create size- and site-controlled quantum structures on demand in GaAsN-based NWs. There are two ways to achieve this: First, parts of the NW can be masked during hydrogenation, as illustrated in \textbf{Figure \ref{fig05}}(a). While the GaAsN bandgap in the unmasked part of the NW is shifted to higher energies during hydrogenation, confinement is created in the protected low bandgap GaAsN below the mask. This technique has been successfully used to create size and site-controlled quantum dots in planar GaAsN quantum wells \cite{Felici2020, Birindelli2014} and can be applied also to NWs to create quantum dots or, as in the sketch, quantum rings. For applications where quantum confinement is critical, the N-containing shell can be grown very thin \cite{Nakama2023, Denis2024} which can be ideal for the design of quantum rings. Second, one can use the opposite approach of hydrogenating the entire NW and then locally removing the implanted H to create the low bandgap structure in which quantum confinement takes place. This can be achieved for example by local laser annealing as illustrated in Figure \ref{fig05}(b). Figure \ref{fig05}(c) shows a proof of concept with a RT $\upmu$-PL scan along a nanowire before hydrogenation, after hydrogenation and after local laser-annealing of the same hydrogenated NW. The individual spectra are shown in the SI 5. The bandgap is shifted locally on a $\upmu$m-scale while optically controlling the process in a $\upmu$-PL setup. A similar approach has been shown to be successful in planar heterostructures \cite{Balakrishnan2012}. The size of the low-bandgap structure is determined by factors such as the size of the laser beam and the heat conduction within the NW. To move from $\upmu$- to nanostructures, the heat source needs to be more localized, which can be achieved by focusing a beam in a dielectric tip or by an electron beam approach, as detailed in previous studies on planar heterostructures \cite{Biccari2018}.

\section{Conclusion}
Hydrogenation of GaAsN-based NWs is proved here to be a versatile technique to tune the bandgap post-growth over a wide energy range between 0.97 - 1.43 eV. Such a wide tuning range can be achieved in the NW geometry due to relaxed strain requirements for the growth of lattice mismatched heterostructures. We show a full shift of the GaAsN bandgap along the polytypic zincblende/wurtzite NWs after hydrogenation, proving N-passivation in the WZ GaAsN, which has not been attempted before. Furthermore, we achieve an order of magnitude increase in optical emission intensity for optimized hydrogenation conditions. For high H-dose, a broad PL emission band is observed between 0.95-1.3 eV, which is due to radiative transitions between a ground state located at Ga-vacancies and a donor level introduced by interstitial H atoms. By thermal annealing, the H implantation can be reversed and the GaAsN bandgap can be controllably tuned up to the value of GaAs. Finally, local bandgap tuning is achieved through punctual laser annealing of a hydrogenated GaAsN NW.    

This work proves that dilute (In)GaAsN is not only an alternative to the InGaAs alloy to provide the appropriate bandgap for telecom photonics, but hydrogenation of (In)GaAsN NWs provides a route to locally control the bandgap post-growth over a wide energy range. 

In summary, this work presents a novel method, hydrogen-based, for achieving bandgap engineering in NWs. Besides the unprecedented energy tunability, the main advantages of this method is that it is post-growth, scalable, does not require device fabrication, is simple and fast (hydrogenation is a few hours-long process), and robust (no material modification was observed in the NWs in the last few years). Such advanced bandgap engineering opens up new possibilities to embed quantum optical devices in NWs for on-chip applications in the near infrared and has also direct applications for the wavelength-optimized design of solar cells, photodetectors and polarization filters.

\section{Experimental Section}
\threesubsection{Measurement methods}\\
The \textmu-PL spectra on single NWs are measured with a 532 nm solid state laser over a power range of 1-10 $\upmu$W, chosen to minimize NW heating, which could promote H out-diffusion or damage the NW. Due to lower detector efficiencies for lower emission energies, the NWs with 4.2\% N were measured with 50 $\upmu$W. All \textmu-PL measurements are performed at RT, to avoid contribution from localized or defect states and evaluate the bandgap energy.
The single NWs are picked individually with glass needles and transferred onto a gold patterned Si-chip to allow reproducible positioning on a specific point on individual NWs. The Si-chip is placed on a motorized stage, which moves with a precision of 50 nm. The light is focused by a 100x Zeiss objective with NA = 0.75, that is optimized for the near infrared to a diffraction limited spot size of 620 nm. The signal is measured using a 0.5 m long spectrometer with 300 g/mm grating blazed for 1 $\upmu$m. Finally, it is recorded by a liquid nitrogen cooled, back illuminated deep-depletion CCD detector optimized for low etaloning in the near infrared and a liquid nitrogen cooled near infrared InGaAs array detector. Individual spectra are corrected for the overall response of the set-up and the respective detectors, which is obtained using the broad emission of a black-body lamp.
The laser annealing is done in a continuous flow cryostat at $T=6$ K to protect the NW from damage by the intense laser beam. To control the laser annealing, the power is continuously increased up to 1000 $\upmu$W and measured in PL to observe changes in the emission energy.

\threesubsection{NW heterostructures}\\
The investigated GaAs/GaAsN/GaAs heterostructure NWs were grown on n-type Si(111) substrates in a molecular beam epitaxy system (MBE) system. The beam equivalent pressure (BEP) of As was $4 \times 10^{-3}$ Pa and the Ga flux was set to match planar growth rates of 1 ML/s or 0.5 ML/s on planar GaAs(001). First, the GaAs NW core was grown by Ga-assisted vapor-liquid-solid (VLS) approach for 30 min at 580$^\circ$C. Second, the Ga catalyst was crystallized and the Ga flux rate is reduced to 0.5 ML/s for the lateral vapour-solid (VS) growth of the shells. Third, a first GaAs shell is grown for 20 min followed by a second growth interruption during which the temperature is reduced to 500$^\circ$C and the N plasma source ignited. Fourth, a GaAs shell is grown for 5 min, followed by a GaAsN shell for 30 min, followed by the outermost GaAs shell for 30 min. For the growth of the GaAsN shell, the shutter is opened and the source is immediately extinguished afterwards. We grew four different series of GaAs/GaAsN heterostructures with N concentrations of 0\%, 0.6\%, 1.6\% and 4.2\% by controlling the microwave power of the plasma source between 0 and 40 W at fixed gas flow rates of 2.5 sccm. The N concentration, as it is referred to in this work, is determined using the BAC model and the measured RT PL emission energy. The N concentration estimated by XRD measurements on GaAsN epilayers grown under identical growth conditions was 0\%, 0.7\%, 2\% and 3\%. Some differences are to be expected due to different growth dynamics in hexagonal shaped NWs compared to planar structures and due to different growth directions. The structural characterization by TEM shows a GaAs core with a diameter of 170 nm surrounded by a 30-50 nm GaAsN shell and a 30-50 nm thick GaAs outermost layer. There is an additional ring in the core, which forms during the ignition of the N-plasma source and it contains approximately 0.2-0.4\% N \cite{Yukimune2019}. An image of the cross-section of a typical NW is shown in Figure \ref{fig01} (a). To acquire that image, electron transparent specimen for scanning transmission electron microscopy (STEM) imaging were prepared in cross-section geometry using a ZEISS Crossbeam 540 focused ion beam operated at acceleration voltages of 30 kV. A 100 nm thick protective platinum layer was first deposited via focused electron beam induced deposition, followed by a thicker 1.5 um platinum layer via focused ion beam induced deposition. The final polishing was done using 2 kV, 10 pA at +-8° incidence. Subsequent STEM imaging was performed on a JEOL JEM F200 cFEG operated at 200 keV. A probe semi-convergence angle of 26 mrad was set and a camera length of 200 mm, leading to a collection semi-angle of 13.4 mrad for bright-field images.
Previous TEM studies along the NW axis have shown a politypic crystal structure with approximately 80\% of ZB and about 20\% of WZ segments, with predominately ZB towards the bottom of the NW. The growth and structural properties have been thoroughly investigated and further details can be found in refs. \cite{Yukimune2019, Yukimune2018, Yukimune2020}. The single wires are transferred with a micromanipulator on a Si-substrate for optical and structural measurements and H implantation.

\threesubsection{H implantation:}\\
For the H implantation, the Si substrate is heated to 230$^\circ$C (290$^\circ$C in SI 3.3) and the NWs are irradiated with ionized H by a low energy Kaufmann source (100 eV ion beam energy) with typical current ion densities of a few tens of $\upmu$ A/cm$^2$. For thermal annealing the Si substrate with the transferred NWs is glued with silver paint onto a metal plate in a high vacuum annealing furnace at $P=10^{-6}$ mbar. The metal plate is heated to temperatures between 300-380$^\circ$C and the NWs are kept for 1 h at the respective temperature.

\medskip
\textbf{Supporting Information} \par
Supporting Information is available

\medskip
\textbf{Acknowledgements} \par 
 The project was funded by the European Union (ERC starting grant, NANOWHYR, 101042349). Views and opinions expressed are those of the author(s) only and do not necessarily reflect those of the European Union or the European Research Council Executive Agency. Neither the European Union nor the granting authority can be held responsible for them. The project was also funded by the Swiss National Science Foundation Ambizione Grant (Grant Number $PZ00P2179801$).
The authors also acknowledge support from MUR (Ministero dell’Università e della Ricerca) through the PNRR MUR project PE0000023-NQSTI. Furthermore support was received from KAKENHI (Nos. 24K21601, 23H00250) by the Japan Society for the Promotion of Science.
\medskip
\bibliographystyle{MSP}
\bibliography{bib.bib}



\end{document}